\documentstyle[12pt,a4]{article}
\input epsf

\begin{document}

\topmargin 0pt
\oddsidemargin 5mm

\setcounter{page}{1}
\begin{titlepage}
\hfill Preprint YeRPHI-1501(1)-98
\vspace{2cm}
\begin{center}

{\bf
 $R$-Parity Violation and Photon-Neutrino Interactions }\\
\vspace{5mm}
{\large R.A. Alanakyan}\\
\vspace{5mm}
{\em Theoretical Physics Department,
Yerevan Physics Institute,
Alikhanian Brothers St.2,

 Yerevan 375036, Armenia\\}
 {E-mail: alanak @ lx2.yerphi.am\\}
\end{center}

\vspace{5mm}
\centerline{{\bf{Abstract}}}
In the framework of the supersymmetric theories with $R$-parity violation we
consider the process $\gamma \gamma \rightarrow \nu_i \bar{\nu_j}$
and its contribution into energy loss from the stars.Under some conditions
 the studied
 contribution exceeds the standard model contribution to the same process.

\vspace{5mm}
\vfill
\centerline{{\bf{Yerevan Physics Institute}}}
\centerline{{\bf{Yerevan 1998}}}

\end{titlepage}

                 {\bf 1.Introduction}

 As known, netrino pairs radiation plays important role in stars
 cooling.
 The process
\begin{equation}
\label{A1}
\gamma \gamma \rightarrow \nu_i \bar{\nu_j}
 \end{equation}

and its various applications (including the stars cooling) has
been considered within the standard model by many
authors  (Fig.1, diagrams b,c) (see \cite{G} - \cite{DF} and references
therein).

If the mass of neutrino is non-zero, the amplitude of the process (1)
is of order $M=\alpha^2 \frac{m_{\nu}}{m^2_e} \frac{1}{m^2_W}$ .In
case of massless neutrinos the process (1) takes place only in the
second order of $G_F$: $M=O(\alpha^2 \frac{p_{\nu}}{m^4_W})$
where $p_{\nu}$ is the momentum of neutrino.

Here we study the contribution to the process (1) within the
supersymmetric models
with $R$-parity breaking \cite{F}- \cite{D}
 described by the diagrams a of the Fig.1\footnote{The amplitudes of
diagrams with scalar charged leptons exchange are supressed by smallness 
of neutrino masses.}. As we see below, the contribution to the process (1)
from scalar neutrions exchange is of order $M=\alpha h^2
\frac{1}{M^2_k}\frac{1}{m_e}$ and, consequently, at small
neutrino and/or sneutrino
masses and/or large Yucawa couplings $h$ it may exceed the standard contribution.

Using $R$-parity violating interaction :

\begin{equation}
\label{A2}
L=h_{ijk}(
\bar{\nu_i}P_L \nu_j +\bar{l_i}P_Ll_j
)\tilde{\nu_k}
\end{equation}
for gauge invariant amplitude of the process (1) we obtain
\footnote{It must be noted that because scalar neutrinos are the
mixture of the
scalar and pseudoscalar, the amplitude of the process
$\tilde{\nu}\rightarrow \gamma \gamma$ is the difference vof the
amplitudes
of the decays
$H^0 \rightarrow \gamma\gamma$ \cite{B}- \cite{VVZS}, $\pi^0 \rightarrow \gamma\gamma$
\cite{S}, \cite{BJ} with replacements
$m^2_{H,\pi} \rightarrow s$ and $\frac{gm_e}{2m_W}\rightarrow h_{ijk}$.}:
\begin{equation}
\label{A3}
M=
\bar{u(k_1)_i}P_L u(k_2)_jA_{a}(k_1)A_{b}(k_2)
(A_{ij}(k_{2a}k_{1b}-g_{ab}(k_1k_2))+B_{ij}
\epsilon^{abcd}k_{1c}k_{2d})
\end{equation}
where
\begin{equation}
\label{A4}
A_{ij}= \frac{\alpha m_e}{\pi} \sum\frac{h_{eek}h_{ijk}}{M_k^2}
\int \limits_{0}^{1} dx \int \limits_{0}^{1-x}dy \frac{1-4xy}{m^2_{e}-i
\epsilon-sxy}.
\end{equation}
\begin{equation}
\label{A5}
B_{ij}=\frac{\alpha m_e}{\pi} \sum \frac{h_{eek}h_{ijk}}{M_k^2}
\int \limits_{0}^{1} dx \int \limits_{0}^{1-x}dy \frac{1}{m^2_{e}-i
\epsilon-sxy},
\end{equation}
$M_k$ is the  mass of the $k$-th generation scalar neutrino.

In the limit $m^2_e \ll s$ integrals in (3),(4) are reduced:
\begin{equation}
\label{A6}
A_{ij}=\frac{2}{3}B_{ij}=\frac{\alpha}{3\pi m_e} \sum \frac{h_{eek}h_{ijk}}{M_k^2}
\end{equation}
The total amplitude of the process (1) is the sum of the
amplitude (3) and standard
amplitude:
\begin{equation}
\label{A7}
M=
B_{SM} \bar{u(k_1)_i}\gamma_5 u(k_2)_jA_{a}(k_1)A_{b}(k_2)
\epsilon^{abcd}k_{1c}k_{2d}
\end{equation}
where in the limit $m_e \ll T$:
\begin{equation}
\label{A8}
B_{SM}= \frac{\sqrt{2} \alpha G_F m_{\nu}}{12 \pi m^2_e}.
\end{equation}
For the cross section of the process (1) we obtain:
\begin{equation}
\label{A9}
\sigma(\gamma \gamma\rightarrow \nu_i \bar{\nu_j})= \frac{ 1}{512 \pi}\frac{1}{\omega_1
\omega_2}(\mid B_{ij} \mid ^2+
\mid B_{ij}-2B_{SM} \mid^2+2\mid A_{ij}\mid ^2)s^3
\end{equation}
where $\omega_{1,2}$ are energies of colliding photons.

For energy loss due to the reaction (1) in the unit of volume and time  we obtain:
\begin{eqnarray}
\label{A10}
&&Q=
 \frac{ 1}{2} \int 
 \frac{ d^3 k_1}{(2 \pi)^3} \frac{2}{e^{\frac{k_1}{T}}-1}
\int \frac{ d^3 k_2}{(2 \pi)^3} \frac{2}{e^{\frac{k_2}{T}}-1}
 (k_1+k_2) \sigma(\gamma \gamma\rightarrow \nu_i \bar{\nu_j})= \nonumber\\
&& = \frac{4!5!\xi(5)\xi(6)}{32 \pi^5}(\mid B_{ij} \mid ^2+
\mid B_{ij}-2B_{SM}\mid^2+2\mid A_{ij}\mid ^2)T^{11}
\end{eqnarray}
The ratio between the sum of the $R$-parity violating+standard
contributions to the process (1) and the pure standard
contribution in (
the absence of $R$-parity violation) is expressed in the following
way:
\begin{equation}
\label{A11}
\frac{Q}{Q_{SM}}=\frac{\mid B \mid ^2+
\mid B-2B_{SM} \mid^2+2\mid A \mid ^2}{\mid B_{SM}\mid ^2},
\end{equation}

In the range where standard contribution may be neglected we
 obtain:
\begin{equation}
\label{A12}
\frac{Q}{Q_{SM}}=\frac{26}{ \pi^2}\sin^4 \theta_W
\frac{h_{eee}^4}{\alpha^2}
\frac{m_e^2}{m^2_{\nu}}\frac{M_W^4}{M^4_k}
\end{equation}

(in formula (12) we suppose that
 the scalar neutrino only of the one flavour contributes).

For instance, at $m_W=M_k$,
$m_e=500 m_{\nu}$, $h_{ee}=10^{-2}$, we have  $Q/Q_{SM}=6.8$.

In \cite{DF} it was shown that standard contribution
from the process (1) to the energy losses in stars at not very large densities
and in some intermediate range of temperatures (at $T< m_e$)
can exceed
 energy losses in stars by neutrino pairs radiation by other mechanisms.
 Consequently, it is actual for
the contribution to the energy losses of the process (1) under study,
at least where it exceeds the standard contribution.

All other applications of the vertex $\gamma \gamma \nu_i
\bar{\nu_j}$
 which has been considered in \cite{DF}
became true also for the contribution considered in this paper ,
because the structure of amplitude (3) is the similar to the
the amplitudes of ref. \cite{DF}.

For example, there may be enhanced the process
$\nu \gamma \rightarrow \nu \gamma $  related to the
 process (1), which responsible for helicity flips of the
 neutrino, which is important for cosmological
 applications (see \cite{DF} and ref. therein).

\newpage

\begin{figure}
\begin{center}
\epsfxsize=10.cm
\leavevmode\epsfbox{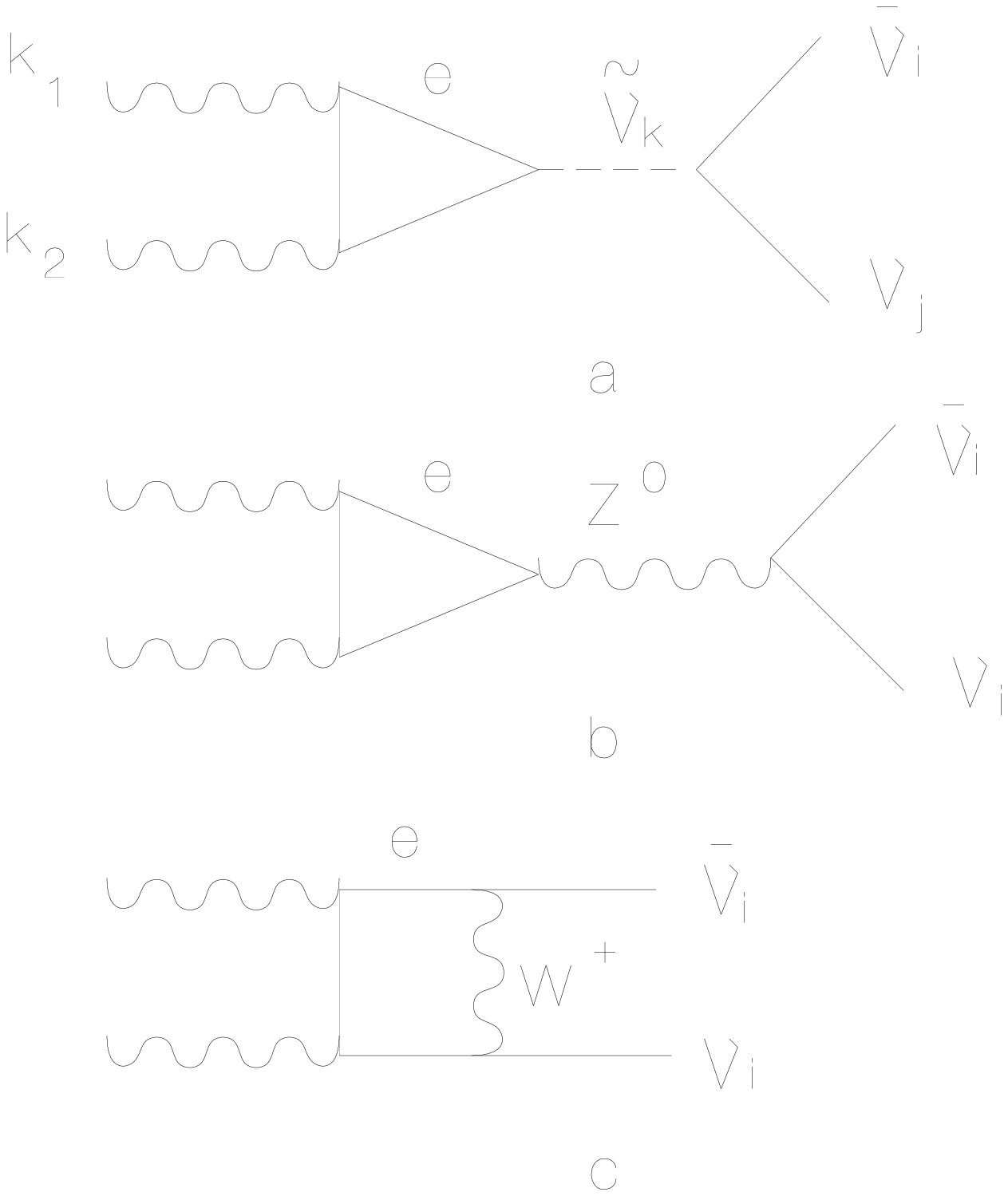}
\end{center}
\caption{
Diagrams corresponding to the processes (1).
}
\end{figure}

\end{document}